\documentclass[twocolumn,pre,aps,showpacs]{revtex4}

\usepackage{subfigure}
\usepackage{epsfig}
\usepackage{bbold}
\usepackage{amsmath}

\def\be{\begin{eqnarray}}
\def\ee{\end{eqnarray}}
\def\ben{\begin{eqnarray*}}
\def\een{\end{eqnarray*}}
\def\bes{\begin{subequations}}
\def\ees{\end{subequations}}

\begin{document}

\title{Variational Formulation of Time-Dependent Density Functional Theory}
\author{J\'er\^ome \surname{Daligault}}
\email{daligaul@lanl.gov}
\affiliation{Theoretical Division, Los Alamos National Laboratory, Los Alamos, NM 87545}

\begin{abstract}
We present a variational formulation of Time-Dependent Density Functional Theory similar to the constrained-search variational formulation of ground-state density-function theory.
The formulation is applied to justify the time-dependent Kohn-Sham method.
Other promising applications to advance TDDFT are suggested.
\end{abstract}


\date{\today}

\maketitle

Ultrafast diagnostics has boosted the importance of reliable and efficient simulations of quantum many-electron dynamics.
The out-of-equilibrium evolution of systems ranging from small molecules to bulk materials has to be described on the fundamental scale over time intervals that exceed the relaxation time.
The leading candidate for yielding tractable algorithms is the time-dependent generalization of the well-established Density Functional Theory (DFT) \cite{Kohn1999}: the Time-Dependent Density Functional Theory (TDDFT) \cite{TDDFTreferences,Ulrich2012,Marquesetal2012}.
To gain the status of a systematic and controlled description, however, TDDFT requires a firm footing.
Its importance has been highlighted by the number and diversity of TDDFT applications \cite{Ulrich2012,Marquesetal2012} and its potential impact has been heightened by the user-accessibility of numerical simulation tools \cite{octopus2012}.
But despite the continuous advances made since the seminal work of Runge and Gross \cite{RungeGross1984}, the foundations of TDDFT are still not as firmly established as those of DFT.
This allowed some authors to openly cast doubt on the integrity of the approach \cite{ShirmerDreuw2007,ShirmerDreuw2008,Schirmer2010} and to stand by their claim despite rebuttal from peers \cite{Maitraetalcomment2008}.
A rigorous basis must be developed if TDDFT is to become an unequivocal cornerstone of future quantum simulations.

An important question that has remained open to this day is whether TDDFT can be rigorously formulated by means of a variational principle similar to that underlying DFT.
Given the unifying and constructive role played by variational formulations in Physics and, especially, in DFT, one can expect that TDDFT would certainly benefit from such a formulation.
In ground-state DFT \cite{DreizlerGross1990}, the density $n_0({\bf r})$ of a system confined by a static potential $v_0({\bf r})$ can be obtained as the global minimum of the energy functional $E[n]=F[n]+\int{d{\bf r}n({\bf r})v_0({\bf r})}$ by solving
\be
\frac{\delta{E[n]}}{\delta n({\bf r})}\Big|_{n=n_0}=0\,,\label{variationalprincipleDFT}
\ee
where $F[n]$ is universal, i.e. independent of $v_0$,
\be
F[n]&=&\inf_{\hat{D}\to n}{\rm Tr}\left[\hat{D}\left(\hat{T}+\hat{W}\right)\right] \label{LLcs1}\\
&=&\sup_v\left[E[v]-\int{d{\bf r}n({\bf r})v({\bf r})}\right]\,, \label{LLcs2}
\ee
with $\hat{T}$ and $\hat{W}$ the kinetic and electron-electron energy operators.
In Eq.(\ref{LLcs1}), the infinum is searched over all $N$-particle density matrices $\hat{D}$ which yield the prescribed function $n({\bf r})$ via
\be
{\rm Tr}\hat{D}\hat{n}({\bf r})=n({\bf r})\,. \label{cDFT}
\ee
This is known as the Levy-Lieb constrained-search procedure, in honor to Levy and Lieb's contributions to substantiate the original proposal of Hohenberg and Kohn \cite{Levy,Lieb}.
Equivalently, $F[n]$ can be obtained as the Legendre transform (\ref{LLcs2}) of the ground-state energy $E[v]={\rm inf}_{\Psi}\langle \Psi| \hat{H}[v] |\Psi\rangle$ when the latter is considered as a functional of the external potential $v({\bf r})$.

The establishment of a variational principle capable to underlie TDDFT has remained a challenging problem for many years that non only gives rise to some deep problem of analysis akin to those encountered in ground-state DFT \cite{Lieb} but also to difficulties specific to the time-dependent case \cite{RungeGross1984,KohlDreizler,GhoshDhara1988,Fukudaetal1994,Rajagopal1996,vanLeeuwen1998,Vignale2008,CohenWasserman2005}.
It was realized that the initial proposal of Runge and Gross \cite{RungeGross1984} based on the Dirac-Frenkel action principle is inadequate since it is inconsistent with causality requirements.
The problem arises since the Dirac-Frenkel principle wrongly fixes boundary conditions at both end points of the action functional, which is incompatible with the first-order character of the Schr{\"o}dinger equation \cite{noteonVignale}.
The difficulty is not unlike that encountered in quantum many-body physics and field theories to extend ground-state perturbation theory to non-equilibrium states.
There, a general and widely used solution was developed by Schwinger and Keldysh, and consists in extending the dynamics along a closed contour in time \cite{Rammer2007}.
Using this artifice, van Leeuwen \cite{vanLeeuwen1998} (and also \cite{Fukudaetal1994}) suggested a route to define an variational action principle that respects causality, but whose steps still remain to be justified.

In this letter, we present a rigorous constrained-search variational formulation of TDDFT that extends the Levy-Lieb formulation of ground-state DFT to the time-dependent case. 
The TD particle density can be obtained as the stationary point of an effective action functional, which is itself defined from a constrained variation of the quantum action of Balian and Veneroni \cite{BalianVeneroni} (in place of the Frenkel-Dirac action); the approach builds on ideas originally developed in field theory \cite{JackiwKerman1979,Eboli1988}.
In the spirit of Levy-Lieb, a universal action functional is defined and is related to the functional introduced by van Leeuwen \cite{vanLeeuwen1998}.
The formulation is then used to justify the TD Kohn-Sham scheme.
For clarity, we successively define our notations, state the main results and derive them.

{\it Definitions.} We are interested in the dynamics over the time interval $t_0\leq t\leq t_f$ of a system of $N$ electrons under the influence of an external, time-dependent scalar potential $v_{0}({\bf r},t)$; we assume that the system evolves from a fixed, known initial state at $t_0$ described by the density operator $\hat{D}_0$; the particle density is denoted by $n_0({\bf r},t)$.

To circumvent difficulties with causality, we regard the dynamics under investigation as a special case of a family of dynamics (i) along a closed-time contour $t(\tau)$ on ${\cal{C}}=[\tau_i,\tau_f]$ that monotonically goes from $t_0$ to $t_f$ and monotonically back to $t_0$ as $\tau$ runs from $\tau_i$ to $\tau_f$, and (ii) governed by the Liouville equation,
\be
i\hbar \frac{d\hat D(\tau)}{d\tau}&\!=\!&t^\prime(\tau)\!\left[\hat{H}_u(\tau),D(\tau)\right]\,\text{with }\hat{D}(\tau_i)=\hat{D}_0\,, \label{Liouvilleequation}
\ee
where $\hat{H}_u(\tau)\!\equiv\!\hat{T}\!+\!\hat{W}\!+\!\int{\!d{\bf r}\hat{n}({\bf r}) u({\bf r},\tau)}$ and $u({\bf r},\tau)$ is an external potential  on ${\cal{C}}$.
With $u\!=\!u_0({\bf r},\tau)\equiv\!v_0({\bf r},t(\tau))$, $u$ is equal on both branches of ${\cal{C}}$, and the dynamics governed by Eq.(\ref{Liouvilleequation}) follows that of the system under investigation along the forward branch and retraces backward the same trajectory along the return branch; therefore, the particle density is also equal on both branches to the interacting density $n_0({\bf r},t(\tau))$.
In general, however, different potentials may act on each branch and, in turn, the forward and backward evolutions differ.
For later reference, we define the evolution operator $\hat{U}_{u}(\tau)=T_Ke^{-\frac{i}{\hbar}\int_{\tau_i}^\tau{d\bar{\tau}t^\prime(\bar{\tau})\hat{H}_{u}({\bar{\tau}})}}$ where $T_K$ is time-ordered chronological operator along the closed-time contour; we also introduce the Heisenberg representation of the particle density along ${\cal{C}}$, $\hat{n}_u ({\bf r},\tau)=\hat{U}_{u}^{-1}(\tau)\hat{n}({\bf r}) \hat{U}_{u} (\tau)$; thus $n_0({\bf r},t(\tau))={\rm Tr}\hat{D}_0\hat{n}_{u_0}({\bf r},\tau)$.

The Liouville equation (\ref{Liouvilleequation}) can be derived by extremizing the extended Balian-Veneroni functional \cite{BalianVeneroni},
\begin{widetext}
\bes
\be
{\cal{A}}_u(\hat{{\cal{D}}},\hat{{\cal{O}}})\!\!&=&\!\int_{\tau_i}^{\tau_f}{d\tau\,{\rm {Tr}}\,\hat{{\cal{O}}}_\tau\left(i\hbar\frac{d\hat{{\cal{D}}_\tau}}{d\tau}-t^\prime(\tau)\left[\hat{H}_u(\tau),\hat{{\cal{D}}}_\tau\right]\right)}-i\hbar\left({\rm Tr}\,\hat{{\cal{O}}}_{\tau_f}\hat{{\cal{D}}}_{\tau_f}-1\right)\\
&=&-\!\int_{\tau_i}^{\tau_f}{d\tau\,{\rm {Tr}}\,\hat{{\cal{D}}}_\tau\left(i\hbar\frac{d\hat{{\cal{O}}_\tau}}{d\tau}-t^\prime(\tau)\left[\hat{H}_u(\tau),\hat{{\cal{O}}}_\tau\right]\right)}-i\hbar\left({\rm Tr}\,\hat{{\cal{O}}}_{\tau_i}\hat{{\cal{D}}}_{\tau_i}-1\right)
\ee
\ees
\label{BalianVeneronivariationalprinciple}
\end{widetext}
under arbitrary variations of the operators (i.e. matrix elements) $\hat{\cal{D}}_\tau$ and $\hat{\cal{O}}_\tau$, and subject to the boundary conditions
\bes
\be
\hat{\cal{D}}_{\tau_i}&=&\hat{D}_0 \label{bc1}\\
\hat{\cal{O}}_{\tau_f}&=&\mathbb{1}\quad\text{(identity operator)}. \label{bc2}
\ee
\label{boundaryconditions}
\ees
For our purposes, we have extended the original Balian-Veneroni action functional along the closed-time contour and added an insignificant constant $i\hbar$ to set its stationary value to zero.
Demanding ${\cal{A}}_u(\hat{{\cal{D}}},\hat{{\cal{O}}})$ to be stationary against arbitrary variations of $\hat{\cal{D}}_\tau$ and $\hat{\cal{O}}_\tau$ not only generates Eq.(\ref{Liouvilleequation}) for $\hat{D}(\tau)$ but also a Liouville equation for $\hat{O}(\tau)$ with boundary conditions (\ref{bc2}) at $\tau_f$.
However, the latter selects the constant solution $\hat{O}(\tau)=\mathbb{1}$ for all $\tau\in{\cal{C}}$ and one is left with Eq.(\ref{Liouvilleequation}) only.
Nevertheless the apparently superflous variational parameter $\hat{\cal{O}}_\tau$ in Eq.(\ref{BalianVeneronivariationalprinciple}) is necessary to ensure that only one boundary condition be associated with the Liouville equation (\ref{Liouvilleequation}), in conformity with its first-order character and thus with causality requirements.
As for the closed-time contour, it ensures causality of the effective action $\Gamma[n]$ defined below.

{\it Main results.} We separate them into four propositions.
For ease of comparison, Table \ref{DFTvdTDDFT} gives the correspondence between the Levy-Lieb formulation of DFT and the present extension to TDDFT.
Below, $\int{\!d\tau}\equiv\int{d\tau\/t^\prime(\tau)}$.

\begin{table}[t]
\begin{tabular}{c||c|c|c|c|c|c}
DFT      & \hspace*{.1cm}$E[n]$\hspace*{.1cm}& \hspace*{.1cm}$E[v]$\hspace*{.1cm} & \hspace*{.1cm}$F[n]$\hspace*{.1cm} & \hspace*{.1cm} Eq.(\ref{variationalprincipleDFT}) \hspace*{.1cm}   & \hspace*{.1cm} Eq.(\ref{LLcs2}) \hspace*{.1cm}  & Eq.(\ref{cDFT}) \\\hline
TDDFT  & $\Gamma[n]$  & $\tilde{A}[v]$ & $F[n]$ & Eq.(\ref{stationarypointofGamma}) & Eq.(\ref{FinTDDFT}) & Eq.(\ref{c1})
\end{tabular}
\caption{Correspondence between DFT and TDDFT. \label{DFTvdTDDFT}}
\end{table}

\noindent {\bf 1-} {\it Constrained-Search Variational Principle:}
Consider the {\it efffective action} functional
\be
\Gamma[n]&=&\underset{\hat{\cal{D}},\hat{\cal{O}}\to n}{\rm s.v.}{\cal{A}}_{u_0}(\hat{\cal{D}},\hat{\cal{O}}) \label{effectiveaction}
\ee
defined for all $n\in{\cal{N}}$ (defined below) as the stationary value (``s.v.'') of ${\cal{A}}_{u_0}$ when $\hat{\cal{D}}_\tau$ and $\hat{\cal{O}}_\tau$ are varied subject (i) to the constraint of density,
\be
\frac{1}{2}{\rm Tr}\left[\left\{\hat{\cal{D}}_\tau,\hat{{\cal{O}}}_\tau\right\}_+\hat{n}({\bf r})\right]&=&n({\bf r},\tau) \label{c1}\,,
\ee
(which upon integration implies ${\rm Tr}\hat{{\cal{D}}}_\tau\hat{{\cal{O}}}_\tau=1$ ), (ii) to the contraint of phase,
\be
{\rm Tr}\hat{\cal{D}}_\tau=1\quad,\quad{\rm Tr}\hat{\cal{O}}_\tau=1\,, \label{c2}
\ee
and (iii) to the boundary conditions (\ref{boundaryconditions}).
${\cal{N}}$ is the set of generalized particle densities consisting of functions $n({\bf r},\tau)$ defined on $\mathbb{R}^3\times {\cal{C}}$ that can be expressed as in Eq.(\ref{c1}) and such that $n({\bf r},\tau_i)=n_0({\bf r},t_0)$ and $\int{d{\bf r} n({\bf r},\tau)}=N$.
Then, the TD density $n_0$ of the system under investigation is a stationary point of $\Gamma[n]$,
\be
\frac{\delta \Gamma[n]}{\delta n({\bf r},\tau)}\Big|_{n=n_0}=0\,. \label{stationarypointofGamma}
\ee
The formulation involves a two-stage, constrained search akin to the Levy-Lieb procedure: first an effective action $\Gamma[n]$ is constructed from a constrained variation of an action, which is then stationarized to obtain $n_0$.
However, while in DFT the stationary value corresponds to a global minimum, it is difficult to characterize further the stationary point obtained using $\Gamma[n]$ \cite{noteonminimization}.
That should not be a worry since problems of motion are generally not influenced by the specific extremum conditions; for instance, in classical mechanics, the least action principle is a stationary principle and the additional criteria of a true extremum are of interest only if stability is involved.

\noindent{\bf 2-} {\it Dual Representation and Universal Effective Action Functional.}
The effective action satisfies
\be
\Gamma[n]=F[n]-\int_{\tau_i}^{\tau_f}{d\tau\int{ d{\bf r} u_0({\bf r},t(\tau)) n({\bf r},\tau)}}
\ee
in terms of the universal (independent of $u_0$) functional
\be
F[n]&\!=\!&\underset{u}{\rm s.v.}\left[-\tilde{A}[u]+\int_{\tau_i}^{\tau_f}{\!d\tau\int{\!d{\bf r} u({\bf r},\tau) n({\bf r},\tau)}}\right]  \label{FinTDDFT}
\ee
where $\tilde{A}[u]={\rm Tr}{\hat D}_0 T_K e^{-\frac{i}{\hbar}\int_{\tau_i}^{\tau_f}{d\tau\hat{H}_u(\tau)}}$ is the van Leeuwen functional introduced in \cite{vanLeeuwen1998} (here extended to mixed states since \cite{vanLeeuwen1998} assumes $\hat{D}_0=|\Psi_0\rangle\langle\Psi_0|$).

\noindent{\bf 3-} {\it Kohn-Sham approach.} The functional $F$ can be decomposed as $F[n]=F_{KS}[n]+F_{ex}[n]$ such that (i) $F_{KS}$ is independent of the interparticle interactions, (ii) the electron-electron interaction contributions are contained in the excess term $F_{ex}$, and (iii) the potential defined by 
\be
%v_{KS}({\bf r},t(\tau))&=&\frac{\delta F_{KS}[n]}{\delta n({\bf r},\tau)}\Big|_{n=n_0}=v_0({\bf r},t(\tau))-\frac{\delta F_{ex}[n]}{\delta n({\bf r},\tau)}\Big|_{n=n_0}
u_{KS}&=&\frac{\delta F_{KS}[n]}{\delta n}\Big|_{n=n_0}=u_0-\frac{\delta F_{ex}[n]}{\delta n}\Big|_{n=n_0} \label{uKS}
\ee
is physical, i.e. equal on both branches, $u_{KS}({\bf r},\tau)\equiv v_{KS}({\bf r},t(\tau))$.
The quantity $\Gamma_{KS}[n]=F_{KS}[n]-\int_{\tau_i}^{\tau_f}{d\tau\int{ d{\bf r} v_{KS}({\bf r},t(\tau)) n({\bf r},\tau)}}$ is the effective action of a noninteracting system of particles in the external potential $v_{KS}$ with the same density $n_0$ as the fully interacting system.
Its dynamics is governed by the single-particle Liouville equation
\be
i\hbar\frac{d\hat\rho_{KS}}{dt}=\left[\hat{p}^2/2m+v_{KS}({\bf r},t),\rho_{KS}\right]\,, \label{KSLiouvilleequation}
\ee
so that $n_0({\bf r},t)={\rm Tr}\rho_{KS}(t)\hat{n}({\bf r})$.

\noindent{\bf 4-} {\it Static limit:} The present formulation can be applied to static external potentials, $v_0({\bf r},t)=v_0({\bf r})$, and equilibrium initial state $\hat{D}_0=e^{-\beta (\hat{T}+\hat{W}+\int{d{\bf r}\hat{n}({\bf r})v_0({\bf r})})}$ at (inverse) temperature $\beta$.
It provides an alternative to the Levy-Lieb formulation of DFT.
The situation is not unlike the duality that occurs in the theory of equilibrium Green's functions between the Keldysh and the finite-temperature (Matsubara) approaches \cite{Rammer2007,GiulianiVignale2005}.

{\it Proofs.}
{\bf 1-}
The proof proceeds by carrying out the s.v. calculation (\ref{effectiveaction}) explicitly using a Lagrange multiplier $J({\bf r},\tau)$ to enforce (\ref{c1}); under independent variation of ${\cal{D}}_\tau$ and ${\cal{O}}_\tau$, two equations emerge,
\bes
\be
i\hbar\frac{1}{t^\prime(\tau)}\frac{d\hat D(\tau)}{d\tau}-\left[\hat{H}_{u_0}(\tau),D(\tau)\right]-\int{d{\bf r}J({\bf r},\tau)\frac{1}{2}\left\{\hat{D}(\tau),\hat{n}({\bf r})\right\}_+}\nonumber\\
=0\quad\text{with }\hat{D}({\tau_i})=\hat{D}_i\hspace*{1cm} \label{cLeqD}\\
i\hbar\frac{1}{t^\prime(\tau)}\frac{d\hat O(\tau)}{d\tau}-\left[\hat{H}_{u_0}(\tau),O(\tau)\right]+\int{d{\bf r}J({\bf r},\tau)\frac{1}{2}\left\{\hat{O}(\tau),\hat{n}({\bf r})\right\}_+}\nonumber\\
=0\quad\text{with }\hat{O}({\tau_f})=\hat{I}\,.\hspace*{1cm} \label{cLeqO}
\ee
\label{extendedLiouville}
\ees
These can be regarded as constrained Liouville equations, where the free dynamics is subjected to an additional non-mechanical ``force'' that maintains the auxiliary condition (\ref{c1}) at all times along ${\cal{C}}$.
In what follows, we first derive a set of relations implied by (\ref{extendedLiouville}) and then apply them.
As we shall see, at that stage, the Lagrange parameter is defined up to a purely time-dependent constant; to keep track of it, we replace $J({\bf r},\tau)$ by $J({\bf r},\tau)+w(\tau)/N$ in Eq.(\ref{extendedLiouville}).
Then as may be verified be direct substitution \cite{noteondetails}, the solution of Eq.(\ref{extendedLiouville}) is
\be
\left\{
\begin{array}{c}
\hat D(\tau)\!=\!\hat{U}_{u_0} (\tau)\hat{C}(\tau,\tau_i)\hat{D}_0\hat{C}(\tau,\tau_i)\hat{U}_{u_0}^{-1}(\tau)e^{-\frac{i}{\hbar}\int_{\tau_i}^\tau{d\bar{\tau} w(\bar{\tau})}}\\
\hat O(\tau)\!=\!\hat{U}_{u_0} (\tau)\hat{C}(\tau_f,\tau)\hat{C}(\tau_f,\tau) \hat{U}_{u_0}^{-1}(\tau)e^{-\frac{i}{\hbar}\int_{\tau}^{\tau_f}{d\bar{\tau} w(\bar{\tau})}} 
\end{array}
\right.
\label{solutionconstainedLiouville}
\ee
with $\hat{C}(\tau,\tau')=T_K e^{-\frac{i}{2\hbar}\int\limits_{\tau'}^{\tau}{\int{d\bar{\tau} d{\bf r} J({\bf r},\bar{\tau})\hat{n}_{u_0}({\bf r},\bar{\tau})}}}$; we note that $\hat{D}(\tau)$ and $\hat{O}(\tau)$ carry a phase related to $w(\tau)$.
Using Eqs.(\ref{solutionconstainedLiouville}) in ${\rm Tr}\hat{{\cal{D}}}(\tau)\hat{{\cal{O}}}(\tau)=1$ yields
\be
-\int _{\tau_i}^{\tau_f}{d\tau w(\tau)}=i\hbar \ln {\rm Tr}\hat{D}_0 \hat{U}_{I}(\tau_i,\tau_f)\equiv W[J]  \label{expWexpression}
\ee
with $\hat{U}_{I}(\tau,\tau^\prime)=T_K e^{-\frac{i}{\hbar}\int_{\tau^\prime}^{\tau}{d\tau\int{d{\bf r}J({\bf r},\tau)\hat{n}_{u_0}({\bf r},\tau)}}}$.
Moreover, substituting Eqs.(\ref{solutionconstainedLiouville}) in the density constaint (\ref{c1}) gives
\be
n({\bf r},\tau)\!=\!\frac{1}{{\rm Tr}\hat{D}_0 \hat{U}_I(\tau_f,\tau_i)}\,{\rm Tr}{\hat{D}_0 T_K\hat{U}_I(\tau_f,\tau)\hat{n}_{u_0}({\bf r},\tau) \hat{U}_I(\tau,\tau_i)}\,, \nonumber\\\label{n_proof_1}
\ee
which is independent of $w(\tau)$.
Incidentally, the right-hand side of Eq.(\ref{n_proof_1}) is also the first derivative of $W[J]$ defined by Eq.(\ref{expWexpression}), and therefore
\be
n({\bf r},\tau)&=&\frac{\delta W[J]}{\delta J({\bf r},\tau)}\,. \label{nequaldeltaWdeltaJ}
\ee
Finally, using Eqs.(\ref{extendedLiouville}), the s.v. (\ref{effectiveaction}) can then be expressed as
\be
\Gamma[n]=-W[J]+\int_{\tau_i}^{\tau_f}{d\tau\int{d{\bf r} J({\bf r},\tau) n({\bf r},\tau)}}\,. \label{LegendretransformGammaW}
\ee
Using Eq.(\ref{expWexpression}), we find that for all purely TD quantities $w(\tau)$,
\ben
-W[J\!+\!w]+\!\int{\!(J+w)n}=-W[J]+\!\int{\! Jn}=\Gamma[n]\,.
\een
and therefore $\Gamma[n]$ is well defined: it is single-valued despite the degeneracy in the Lagrange multiplier for a given $n$.
The normalization constraint (\ref{c2}) allows one to fix $w(\tau)$ in Eqs.(\ref{solutionconstainedLiouville}) to
\ben
w(\tau)=\frac{{\rm Tr}{\hat{D}_0 T_K \int{d{\bf r}J({\bf r},\tau)\hat{n}_{u_0}({\bf r},\tau) e^{-\frac{i}{\hbar}\int\limits_{\tau_i}^{\tau}{\int{d\bar{\tau} d{\bf r} J({\bf r},\bar{\tau})\hat{n}_{u_0}({\bf r},\bar{\tau})}}}}}}{{\rm Tr}{\hat{D}_0 T_K e^{-\frac{i}{\hbar}\int\limits_{\tau_i}^{\tau}{\int{d\bar{\tau} d{\bf r} J({\bf r},\bar{\tau})\hat{n}_{u_0}({\bf r},\bar{\tau})}}}}}\,,
\een
and therefore to define a one-to-one mapping between $n$ and $J$ (the $J\to n$ mapping is defined through Eq.(\ref{nequaldeltaWdeltaJ})).
In particular, with $n=n_0$, this gauge selects the natural solution $J\equiv 0$ and $w\equiv 0$.
Under these conditions, Eq.(\ref{nequaldeltaWdeltaJ}) says that $n$ and $J$ are conjugate variables and the relation (\ref{LegendretransformGammaW}) for $\Gamma[n]$ can be regarded as the Legendre transform of $W[J]$; it follows that
\be
\frac{\delta{\Gamma[n]}}{\delta n ({\bf r},\tau)}=J({\bf r},\tau)\,. \label{deltagammadeltan}
\ee
The removal of constraints, which corresponds to the physical situation under consideration, is equivalent to $J=0$; hence, from Eq.(\ref{deltagammadeltan}), $\Gamma[n]$ is stationary with respect to the variation of $n$ around $n_0$.

{\bf 2-} Using $\hat{U}_{u_0+J}(\tau_f,\tau_i)=\hat{U}_{u_0}(\tau_f,\tau_i)\hat{U}_I(\tau_i,\tau_f)$ and $\hat{U}_{u_0}(\tau_f,\tau_i)=\mathbb{1}$, we find 
\ben
W[J]&=&\tilde{A}[u^\star= u_0+J]\,.
\een
Using Eq.(\ref{LegendretransformGammaW}), it follows (dropping variables)
\ben
\Gamma[n]&=&-\tilde{A}[u^\star]+\int{nJ}\\
&=&\left[-\tilde{A}[u^\star]+\int{n u^\star}\right]-\int{u_0 J}\\
&=&\underset{u}{\rm s.v.}\left[-\tilde{A}[u]+\int_{\tau_i}^{\tau_f}{\!d\tau\int{\!d{\bf r} u({\bf r},\tau) n({\bf r},\tau)}}\right]-\int{u_0 J}\\
&=&F[n]-\int{u_0 J}\,,
\een
which proves proposition 2.

{\bf 3-}
A complete proof is beyond the scope of the paper, and we hope to publish it elsewhere.
We outline the main elements and point to relevant references for details.
By treating the strength of the electronic Coulomb repulsion $g=e^2$ as an expansion parameter, the effective action $F[n]$ can be expressed as
\be
F[n]=\sum_{k=0}^\infty {g^k F^{(k)}[n]}\,. \label{e2expansion}
\ee
Different techniques can be used to obtain the coupling-constant expansion (\ref{e2expansion}) from Eq. (\ref{FinTDDFT}), such as the auxiliary field method \cite{Fukudaetal1994,NegeleOrland1988} or the inversion method \cite{Fukudaetal1994,Okumara1996,ValievFernando1997}.
The zeroth-order contribution $F_{KS}\equiv F^{(0)}$, which is the only remaining term in the limit $g\to 0$, corresponds to the universal effective action of a non-interacting system, while the higher-order corrections collected in $F_{ex}\equiv\sum_{k=1}^\infty {g^k F^{(k)}[n]}$ contain the effect of electron-electron interactions, including the TD Hartree and KS exchange contributions at lowest order.
Then, Eq. (\ref{uKS}) is obtained by combining $\delta\Gamma/\delta n|_{n_0}=\delta F/\delta n|_{n_0}-u_0=0$ with $F=F_{KS}+F_{ex}$.
From the symmetry property of the effective action under exchanging the forward and backward components of $n$ along ${\cal{C}}$, the solution of $\delta F_{KS}/\delta n$ evaluated at the physical density $n_0$ can be shown to be equal on both branches.
Finally, from propositions 1 and 2, $\Gamma_{KS}[n]=F_{KS}[n]-\int{v_{KS} n}$ can be interpreted as the effective action $\displaystyle \underset{\hat{\cal{D}},\hat{\cal{O}}\to n}{\rm s.v.}{\cal{A}}_{u_{KS}}(\hat{\cal{D}},\hat{\cal{O}})$.
At $n=n_0$, the stationary point implies the free Liouville equation, which, for a system of independent particles, is equivalent to the single-particle Liouville equation (\ref{KSLiouvilleequation}) for the one-particle density operator $\hat\rho_{KS}$.
For a pure initial state, the latter can be diagonalized and one obtains the more usual single-particle KS equations for the KS orbitals.
We remark that, in the absence of a variational principle, the TDKS scheme was either justified by assuming the so-called non-interacting $v$-representability of interacting densities, or searched by direct construction of the KS system through solution of a non-trivial partial-differential equation for $v_{KS}$ \cite{Maitraetal2010,vanLeeuwen1999}.

In conclusion we have described a constrained-search variational formulation of TDDFT.
The key results are collected in the four propositions listed above and the main ingredients are conveniently compared with those of the Levy-Lieb constrained-search formulation in Table (\ref{DFTvdTDDFT}). 
The approach integrates and extends ideas previously developed in other fields \cite{vanLeeuwen1998,Fukudaetal1994,BalianVeneroni,Eboli1988}.
A profound mathematical analysis of its underpinnings, well beyond the scope of the present paper, would be much desirable; we hope that the incompleteness of the results presented here will encourage others to pursue some of the questions raised by them.
We anticipate that many of the techniques developed in field theory, e.g. functional integration, loop expansion, auxiliary field..., could be much valuable to develop controlled approximations based on $\Gamma[n]$ and to provide additional insights on the TD Kohn-Sham potential.
Finally, the approach can be expanded to TD-current-DFT \cite{Ulrich2012} and can also be adapted to purely classical systems \cite{ChanFinken2005} by considering the classical counterparts of the different quantities (e.g., density operators into distribution functions, commutator into Poisson bracket...)

This work was carried out under the auspices of the National Nuclear Security Administration of the U.S. Department of Energy at Los Alamos National Laboratory under Contract No. DE-AC52-06NA25396.


\begin{references}
\bibitem{noteonVignale} More recently, Vignale \cite{Vignale2008} showed that causality is respected if the final endpoint is not constrained, but at the additional price of losing the stationarity of the action functional at the physical density.
\bibitem{noteonminimization} It may nevertheless be possible to show that the physical density corresponds to a minimum of the convex, euclidean effective action obtained from $\Gamma[n]$ by continuing time to imaginary time in the formalism $t\to it$.
\end{references}
\end{document}